\documentclass[iop]{emulateapj}

\newcommand{\avgm}{\langle M \rangle}

\shorttitle{X-RAY AND OPTICAL MICROLENSING IN Q~J0158--4325}
\shortauthors{MORGAN ET AL.}

\begin{document}

\title{Further Evidence that Quasar X-Ray Emitting Regions Are Compact: X-Ray and Optical Microlensing in the Lensed Quasar Q~J0158--4325\altaffilmark{1}}

\author{Christopher~W.~Morgan\altaffilmark{2}, Laura~J.~Hainline\altaffilmark{2}, Bin~Chen\altaffilmark{3}, Malte~Tewes\altaffilmark{4}, Christopher~S.~Kochanek\altaffilmark{5}, Xinyu~Dai\altaffilmark{3}, Szymon~Kozlowski\altaffilmark{5,7}, Jeffrey~A.~Blackburne\altaffilmark{5}, Ana~M.~Mosquera\altaffilmark{5}, G.~Chartas\altaffilmark{6}, F.~Courbin\altaffilmark{4}, \& G.~Meylan\altaffilmark{4}}

\altaffiltext{1}{Support for this work was provided by the National Aeronautics and
Space Administration through Chandra Award Number GO0-11121 issued by
the Chandra X-ray Observatory Center, which is operated by the
Smithsonian Astrophysical Observatory for and on behalf of the National
Aeronautics Space Administration under contract NAS8-03060}

\altaffiltext{2}{Department of Physics, United States Naval Academy, 572C Holloway Road,
Annapolis, MD 21402, cmorgan@usna.edu}

\altaffiltext{3}{Department of Physics and Astronomy, University of Oklahoma, 440 W. Brooks Street, Norman, OK 73019}

\altaffiltext{4}{Laboratoire d'Astrophysique, \'{E}cole Polytechnique F\'{e}d\'{e}rale de Lausanne (EPFL),
Observatoire, 1290 Sauverny, Switzerland}

\altaffiltext{5}{Department of Astronomy, The Ohio State University, 140 West 18th Avenue, Columbus, OH 43210
-1173}

\altaffiltext{6}{Department of Physics and Astronomy, College of Charleston, 58 Coming Street, Charleston, SC 29424}

\altaffiltext{7}{Warsaw University Observatory, Al. Ujazdowskie 4, 00-478, Warszawa, Poland}

\clearpage

\begin{abstract}
We present four new seasons of optical monitoring data and six epochs of X-ray photometry for the doubly-imaged lensed quasar Q~J0158--4325.  
The high-amplitude, short-period microlensing variability
for which this system is known has historically precluded a time delay measurement by conventional methods. We attempt to circumvent this limitation by application
of a Monte Carlo microlensing analysis technique, but we are only able to prove that the delay must have the expected sign (image A leads image B).
Despite our failure to robustly measure the time delay, we successfully model the microlensing at optical and X-ray wavelengths to find a
half light radius for soft X-ray emission 
$\log(r_{1/2,X,soft}/{\rm cm}) = 14.3^{+0.4}_{-0.5}$, an upper limit on the half-light radius for hard X-ray emission $\log(r_{1/2,X,hard}/{\rm cm}) \leq 14.6$
and a refined estimate of the 
inclination-corrected scale radius of the optical $R$-band (rest frame 3100~\AA) continuum emission region of $\log(r_{s}/{\rm cm}) = 15.6\pm0.3$.
 
\end{abstract}


\section{Introduction}

Quasars can be highly variable. Since this variability arises in the quasar continuum source itself, it appears in all the
images of a gravitationally lensed quasar, but at different times in each image due to 
differences in the gravitational potential of the lens galaxy and the geometric length of the paths of  
each image \citep{Refsdal1964,Cooke1975}. 
As first pointed out by \citet{Chang1979}, the stars near each lensed image
also cause temporal variations in the magnification of
each image, leading to additional variability.  This ``microlensing'' variability is not 
correlated between the images, so it functions as both a difficult source of noise in time delay measurements and as a probe of quasar structure because the
size of the emission region ultimately controls the amplitude of the microlensing variability.

\begin{figure*}
\epsscale{1.0}
\plotone{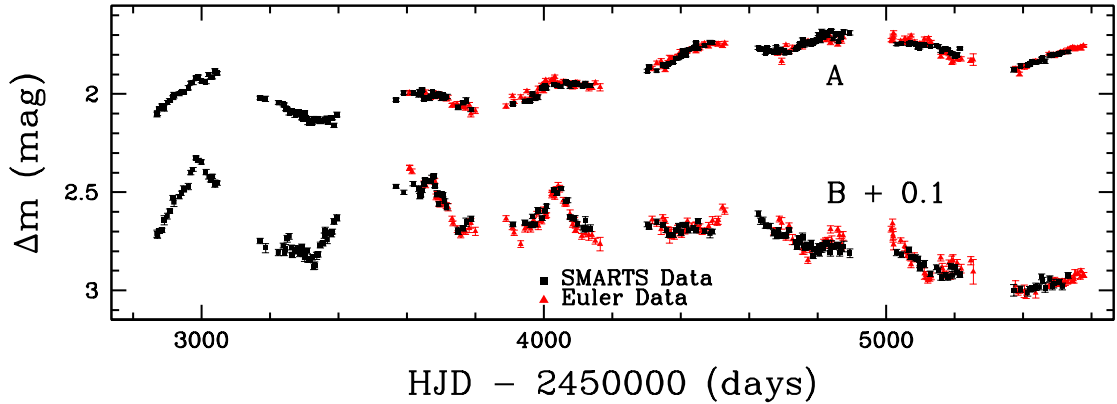}
\caption{Top Panel: QJ0158--4325 $R$-band light curves for the magnitudes of images A 
and B + 0.1. Data from the SMARTS 1.3m are plotted as black squares.  Data from the 1.2m Euler Swiss Telscope are plotted as red triangles. 
Observations failing a seeing limit of ${\rm FWHM} \leq 2\farcs0$ are not shown.  This eliminated 20 points from the light curves.}
\label{fig:lightcurves}
\end{figure*}

Traditional time delay measurements require either that the microlensing variability is of sufficiently small amplitude so as to be insignificant in the analysis or that its timescale
is significantly longer than that of the quasar's intrinsic variability, allowing for some sort of parametric modeling and subtraction
\citep[e.g.][]{Ofek2003,Paraficz2006,Kochanek2006,Poindexter2008}.  Attempting to measure time delays in systems with significant
short-timescale microlensing is both difficult and error prone 
\citep[see][for quantitative characterizations of the influence of microlensing on time delay estimates]{Eigenbrod2005}.
In \citet{Morgan2008}, we applied a new approach to analyzing the light curves of two lensed quasars with significant microlensing and time
variability, in which we attempted to simultaneously model the 
microlensing variability using the Bayesian Monte Carlo technique of \citet{Kochanek2004}.  In our new approach, we 
marginalized over the physically realistic models of the microlensing 
signal to yield a measurement of the most likely time delay.  We successfully confirmed the time delay of
HE~1104--1805 but failed to measure the delay in Q~J0158--4325. In the present paper, we supplement our previously published light curves with 4 new seasons of 
monitoring data from two observatories and make a second attempt.

In the models of \citet{Morgan2008}, we could also marginalize over the time delay uncertainties and use the microlensing to study the structure of the quasar.
The amplitudes and the timescales of the variability due to quasar microlensing are a function of a number of physical variables such as the continuum source size, 
the mean properties of the lens in the vicinity of the lensed images, and 
the relative velocities of the source, the lens and the observer.  During the last decade a variety
of techniques have been developed to analyze the effects of microlensing on lensed quasar images.  Single-epoch photometry in which 
one or more of the quasar image fluxes are anomalous (i.e. deviates significantly from the predictions of the macroscopic lens model or shows significant wavelength 
dependencies) can provide (prior dependent) estimates of source sizes and constrain temperature profiles 
\citep[e.g.][]{Pooley2006,Pooley2007,Bate2008,Floyd2009,Blackburne2011,Mosquera2011,Mediavilla2011}.  These studies must also assume a mean mass for
the microlensing stars.

Stronger and less prior-dependent constraints can be obtained from analyses of quasar microlensing variability.  
We have employed the technique of \citet{Kochanek2004} to analyze uncorrelated variability in
quasar optical lightcurves to yield a correlation between the sizes of quasar accretion disks and the masses of their black holes \citep{Morgan2010}, 
measurements of the structure of quasar accretion disks \citep[e.g.][]{Poindexter2008}, and 
even a constraint on the orientation of a quasar's accretion disk \citep[in Q2237+0305,][]{Poindexter2010}.  We have also modified the \citet{Kochanek2004} technique
to simultaneously analyze optical and X-ray lightcurves, yielding measurements of the size of the X-ray emitting regions in four 
systems, PG1115+080, HE~1104--1805, RXJ~1131--1231                      
and HE~0435--1223 \citep[][respectively]{Morgan2008,Chartas2009,Dai2010,Blackburne2012}.   
In those papers, we found that the X-ray continuum emission emerges from region $\sim10 r_g$ in radius, where $r_g$ is the black hole's gravitational radius 
$r_g = G M_{BH}/c^2$. This is at least an order of magnitude smaller than sizes measured in the UV/optical, consistent with the arguments of
\citet{Pooley2007}.   These results seem to disfavor 
X-ray continuum emission from an extended disk-corona model \citep[e.g.][]{Haardt1991,Merloni2003}, but more measurements are needed both to confirm
these results and to begin searching for correlations between the X-ray emitting regions and other physical properties of the quasars.
In this paper, we add to the sample of quasars studied at X-ray wavelengths by applying our Monte Carlo 
microlensing analysis technique to the combined optical and X-ray lightcurves of QJ0158--4325 to yield size estimates
for the hard and soft X-ray continuum emission regions. Making use of our new
constraints on the time delay, we are also able to significantly refine our estimate of the optical accretion disk size.

In \S~\ref{sec:obs} we begin by describing our {\it Chandra} X-ray observations \citep{Chen2012}, followed by a 
presentation of our new optical monitoring data and a brief review of 
our data reduction process. In \S~\ref{sec:td}, we describe our
time delay measurement attempts and in \S~\ref{sec:results} we present the results from our Monte Carlo microlensing simulations.  
In \S~\ref{sec:conclusions}, we conclude with a discussion of the implications for the physics of the optical and X-ray continuum sources in this quasar.
All calculations in this paper assume a flat $\Lambda$CDM cosmology with $h=0.7$, $\Omega_{M}=0.3$ and $\Omega_{\Lambda}=0.7$.

\section{Observations and Light Curve Reduction}
\label{sec:obs}

We imaged the doubly lensed quasar Q~J0158--4325 \citep{Morgan1999} in X-rays using the ACIS imaging spectrometer \citep{Garmire2003} on the 
{\it Chandra X-Ray Observatory} \citep[{\it Chandra},][]{Weisskopf2002} on six 
occasions between 2009 November 6 and 2010 October 6.  These observations were a component of 
a larger {\it Chandra} Cycle 11 monitoring program,  the details of which are published in \citet{Chen2012}. We divided the full observed frame X-ray 
spectral range (0.4--8.0 keV) into soft (0.4--1.3 keV) and hard (1.3--8.0 keV) energy bands.  We analyze the soft and hard X-ray variability separately in order
to constrain the energy structure of the X-ray continuum source.

\def\hm{\hphantom{-}}
\begin{deluxetable}{ccccc}[h]
\tabletypesize{\scriptsize}
\tablecaption{Q~J0158--4325 Light Curves}
\tablewidth{0pt}
\tablehead{ HJD &\multicolumn{1}{c}{Seeing}
                &\multicolumn{1}{c}{QSO A (mags)} &\multicolumn{1}{c}{QSO B (mags)} 
                &\multicolumn{1}{c}{Source} 
              }
\startdata
$4316.305$ & $1.55$ & $ 1.869\pm 0.012$ & $ 2.541\pm 0.018$ & Euler \\ 
$4329.389$ & $1.34$ & $ 1.881\pm 0.008$ & $ 2.550\pm 0.013$ & SMARTS \\ 
$4329.396$ & $1.61$ & $ 1.842\pm 0.009$ & $ 2.558\pm 0.015$ & Euler \\ 
$4335.299$ & $2.10$ & $( 1.872\pm 0.009)$ & $( 2.567\pm 0.015)$ & SMARTS \\ 
$4342.326$ & $1.31$ & $ 1.833\pm 0.014$ & $ 2.589\pm 0.025$ & Euler \\ 
$4347.368$ & $1.80$ & $ 1.861\pm 0.010$ & $ 2.531\pm 0.016$ & Euler \\ 
$4348.295$ & $1.20$ & $ 1.853\pm 0.008$ & $ 2.565\pm 0.012$ & SMARTS \\ 
$4354.319$ & $2.01$ & $( 1.876\pm 0.010)$ & $( 2.544\pm 0.015)$ & Euler \\ 
$4357.293$ & $1.30$ & $ 1.861\pm 0.008$ & $ 2.593\pm 0.012$ & SMARTS \\ 
\enddata
\tablecomments{HJD is the Heliocentric Julian Day -- 2450000 days.
The QSO A\&B
columns give the magnitudes of the quasar images relative to the
comparison stars. 
A few points in the light curves with seeing FWHM~$\geq 2\farcs0$
(in parentheses) were not used in the analysis.  Table 1 will be published in its entirety in the electronic edition of {\it The Astrophysical
Journal}.  A portion is shown here for guidance regarding its form and content.}
\label{tab:lightcurves}
\end{deluxetable}
 
 \begin{figure*}
\epsscale{1.0}
\plotone{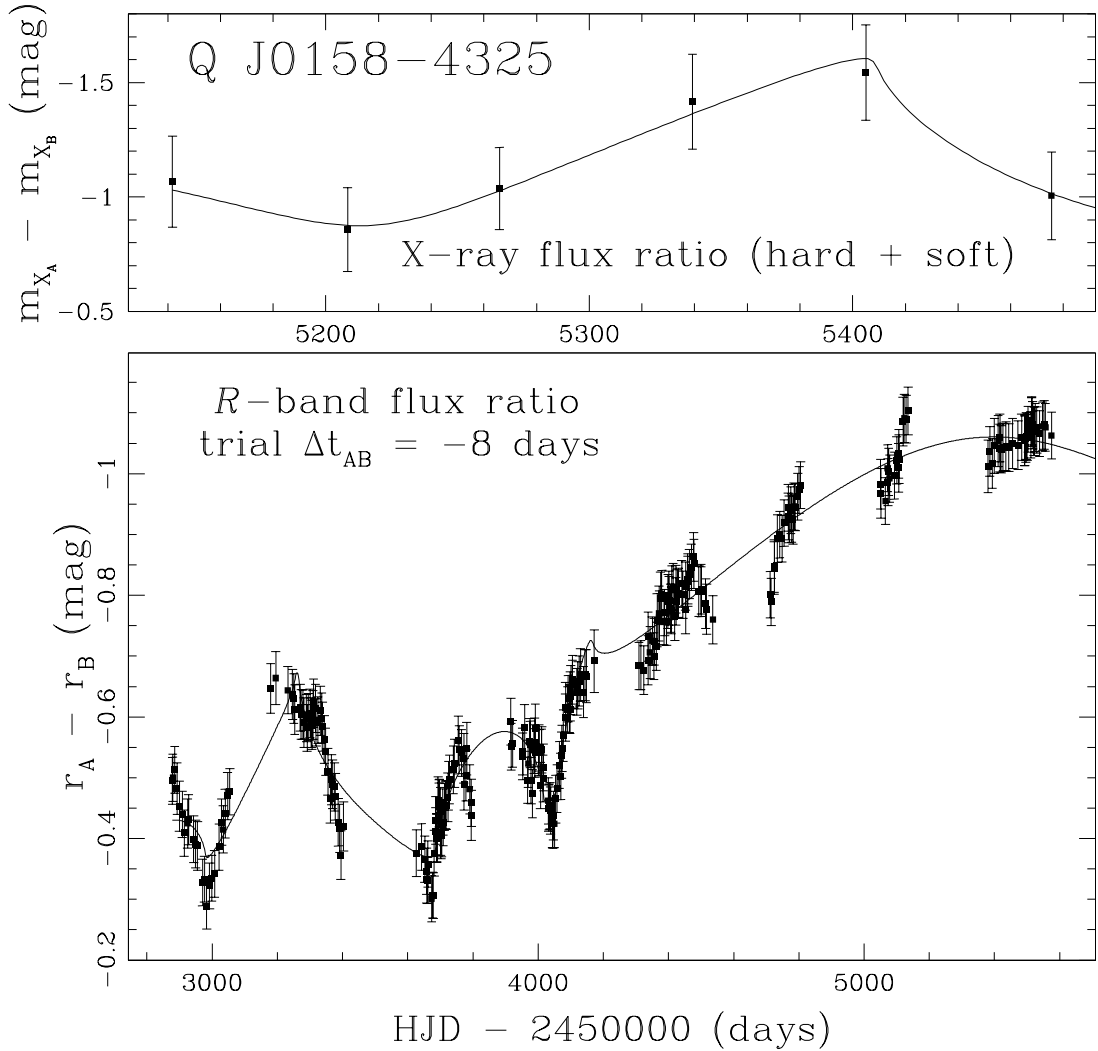}
\caption{Example of Monte Carlo fits to the observed flux ratios in the lensed quasar Q~J0158--4325. 
Top Panel: Six epochs of the observed $A/B$ X-ray flux ratio from our {\it Chandra} monitoring program shown as $m_{X_A} - m_{X_B}$.  
An example of a good Monte Carlo fit to these measurements is also shown.
Bottom Panel: The Q~J0158-4325 optical difference light curve for a trial time delay of $\Delta t_{AB} = - 8$~days.  Data are shown with error bars.  The
curve is a good fit ($\chi^2/N_{dof} \approx 1.2$) from our Monte Carlo analysis. Note its failure to fully match all features of the trial difference light curve. }
\label{fig:fits}
\end{figure*}
   
We have monitored Q~J0158--4325 for eight seasons
in the $R$-band using the SMARTS 1.3m telescope with the  
ANDICAM optical/infrared camera \citep{Depoy2003}\footnote{http://www.astronomy.ohio-state.edu/ANDICAM/} and using the 1.2m Euler 
Swiss Telescope as a part of 
the COSMOGRAIL\footnote{http://www.cosmograil.org/} project. We analyzed the optical images from both observatories simultaneously, employing the deconvolution
algorithm of \citet{Magain1998}.  In this technique, we measure the point spread function (PSF) in each frame using the same 5 reference
stars in the field.  The astrometry of the lens components is held fixed to the 
values measured in Hubble Space Telescope $H$-band images from the CfA-Arizona Space 
Telescope Lens Survey (CASTLES), the details of which were published in \citet{Morgan2008}.  Using the corresponding PSFs, we simultaneously 
fit the photometry of the point sources and a regularized image of the lensing galaxy to all the frames. The resulting QSO image fluxes are calibrated 
relative to the deconvolved field stars. Finally, we apply a small multiplicative and additive correction to minimize the dispersion in the overlapping sections of the 
Euler and SMARTS lightcurves. This empirical adjustment is required to compensate for small color terms introduced by differences in the filters and
quantum efficiency curves of the CCDs. The new photometric data are presented in Table~\ref{tab:lightcurves},
and our composite eight-season light curves are displayed in Figure~\ref{fig:lightcurves}.

\section{Analysis}
\label{sec:td}

\subsection{Lightcurve Preparation}
\label{sec:lc}
 
\begin{figure*}
\epsscale{1.0}
\plottwo{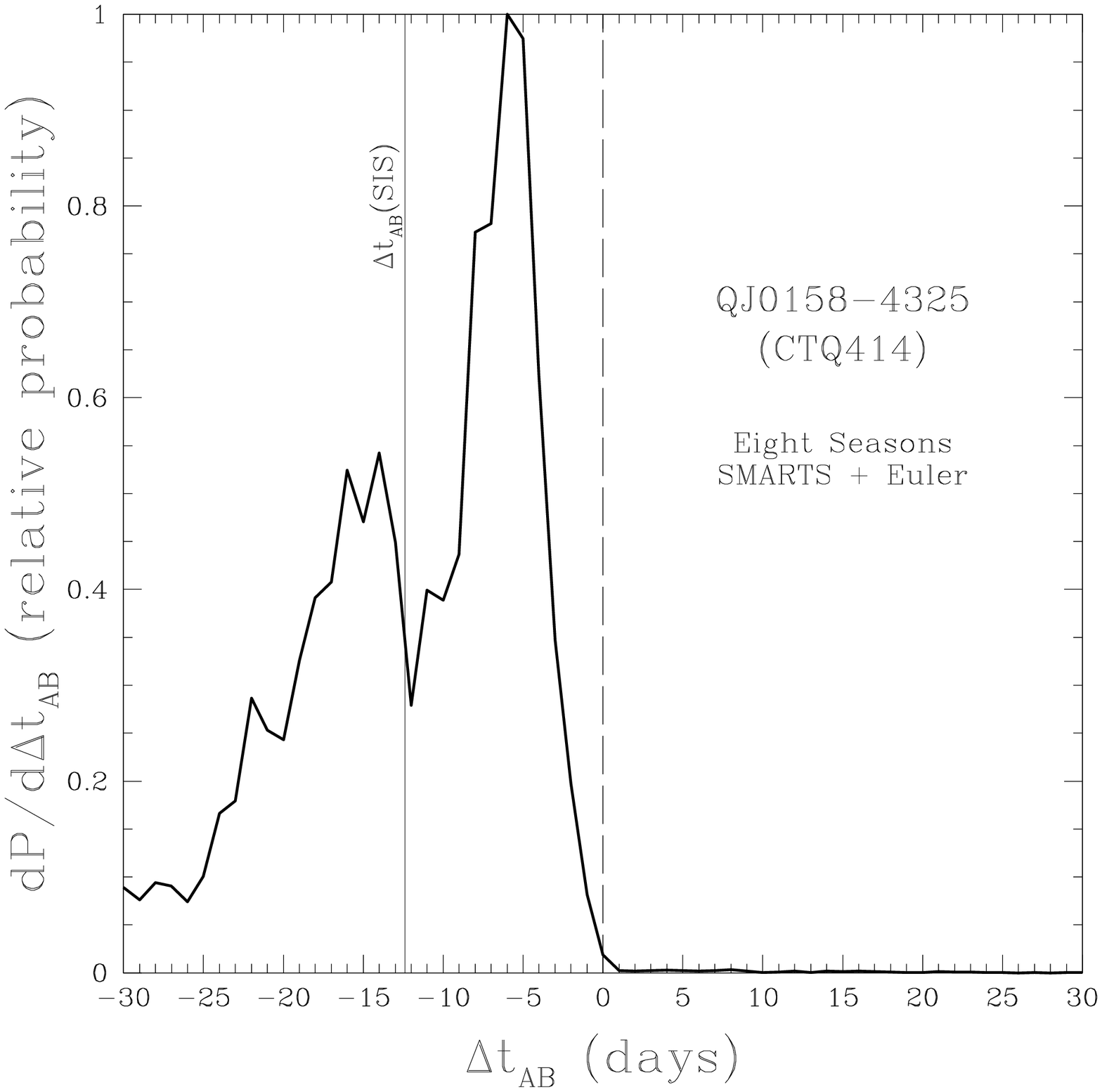}{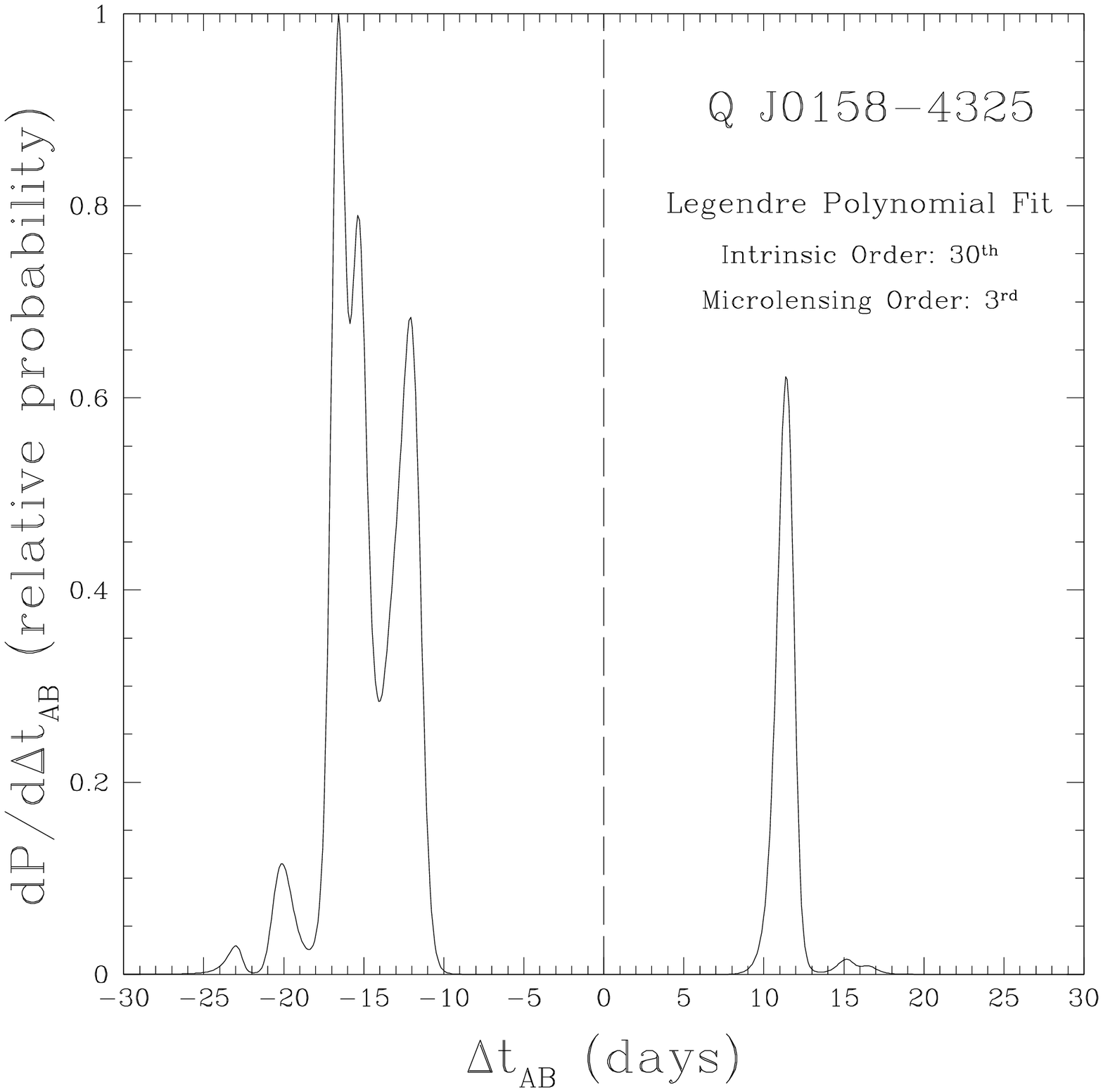}
\caption{Left: Probability distribution for the time delay $\Delta t_{AB} = t_A - t_B$ in Q~J0158--4325, where we restricted the extrapolation of the light curves to be
less than 5 days.  Negative delays are now strongly favored over positive delays as predicted by lens models, but our analysis did not converge on a single value for the delay. 
The solid vertical line at $\Delta t_{AB} = -12.4$~days represents the predicted delay of a SIS lens model. 
Right: Results of Legendre polynomial fits to the observed lightcurves; relative probability as a function of time delay 
$\Delta t_{AB} = t_A - t_B$ is plotted.  The intrinsic and microlensing variability
are fit with 30$^{{\rm th}}$ and 3$^{{\rm rd}}$ order polynomials, respectively.  Note the absence of a single, well-defined peak.}
\label{fig:td}
\end{figure*}

We began our analysis by shifting the image A light curve by a reasonable set of trial time delays ($-30$~days $\leq \Delta t_{AB} = t_A - t_B \leq +30$~days) 
separated by one day intervals, where A leads B for $\Delta t_{AB} < 0$.  The model prediction for a singular isothermal sphere (SIS) lens model is that 
A leads B by $\Delta t_{AB} = -12.4$~days.
We chose to shift image A because it is less variable than image B.  
Our analysis algorithm requires contemporaneous pairs of A/B data points, in most cases necessitating linear interpolation between bracketing epochs. 
For shifted points that fell into the inter-season gaps, we permitted 5 days of linear extrapolation, and we increased the size of the measurement uncertainties 
on the extrapolated points using our estimate of the quasar's intrinsic variability structure function.  
Since the Bayesian integral shown in Eq.~\ref{eq:bayes} is actually a discrete sum in which the 
statistical weight of each trial time delay must be held fixed, all trial lightcurves must have the same number of epochs, forcing us to 
discard some data points for the shorter trial delays. The details of this selection technique are described in \citet{Morgan2008}. 

We also generated a second set of trial light curves, permitting unlimited extrapolation in the inter-season gaps using the Damped Random Walk (DRW) models of
\citet{Kelly2011,Kozlowski2010} and \citet{MacLeod2010}.  Using the implementation of \citet{Kozlowski2010}, we first modeled 
the raw lightcurves, yielding a long term variability structure function measurement 
$\rm{SF_\infty}=0.337$~mags and timescale $\tau = 1867$~days for image A and $\rm{SF_\infty}=0.280$~mags and timescale $\tau = 419.3$~days for image B.
These models also provide the structure function estimate used in our first approach.  
The obvious advantage of this treatment is that no data must be discarded, 
however the DRW process models the intrinsic variability in the light curves of only one image at a time such that the variability in the DRW points is not correlated with 
the contemporary points in the other image.  
In the absence of the strong microlensing signal, we would model the light curves jointly, following the procedure of \citet{Press1992}.
The introduction of additional uncorrelated variability can be significant, particularly for the longer trial delays, 
forcing the models to fit the differences in the DRW interpolations as microlensing. This further blurred any effects of the time delay, so we
do not report on these results.

Because we have only 6 epochs of X-ray observations, we are unable to shift the X-ray 
measurements by the trial time delay, since doing so would eliminate most of our data points.
Both our macroscopic mass models and this present work predict a relatively short time delay $-30 \leq \Delta t_{AB} \leq  0$~days, but it is likely that
some intrinsic variability also contaminates the X-ray flux ratios.  To account for this, we estimated the variability structure function of the 
(sparsely sampled) X-ray light curves and then conservatively assumed a 30~day time delay, yielding an additional systematic uncertainty of 0.1~magnitudes
in the X-ray flux ratios. This additional systematic uncertainty limited the precision of our Monte Carlo analysis, but it cannot be avoided until the time delay is measured.

\subsection{Simultaneous Microlensing and Time Delay Analysis}
\label{sec:micro}

Using the same technique that we described in detail in \citet{Morgan2008}, we applied the Monte Carlo 
microlensing analysis method of \citet{Kochanek2004} to the light curves from each trial time delay, in essence attempting to
reproduce the observed lightcurves by random combinations of the unknown variables: the lens galaxy stellar mass fraction, the median microlens mass
$\avgm$ and the effective velocity between source, lens and observer $\vec{v}_e$. We used the recently measured lens redshift $z_l = 0.317$ \citep{Faure2009}
to update our set of ten lens models, each consisting of concentric NFW \citep{Navarro1996} and de Vaucouleurs mass profiles.
We parametrize the stellar mass fraction $0.1\leq f_{M/L}\leq1$ in our model sequence by comparing the fractional mass of the de Vaucouleurs component 
to the mass of a constant M/L model ($f_{M/L}=1$), varying $f_{M/L}$ between 0.1 and 1.0 in uniform steps.  For each model, we generate four unique sets 
of microlensing magnification patterns using the ${\rm P^3 M}$ method described in \citet{Kochanek2004}. The magnification patterns are $4096 \times 4096$ pixel
arrays with an outer scale radius of $20 R_{E}$, where $R_E = 3.4 \times 10^{16} (\langle M \rangle/ 0.3 M_\sun)^{1/2}$ is the Einstein radius,  
yielding a source plane pixel scale of $1.7 \times 10^{14} (\langle M \rangle/ 0.3 M_\sun)^{1/2}$~cm. We performed four separate 
Monte Carlo simulations, one per set of magnification patterns.  In each simulation, we attempted $10^8$ trials per mass model, and in each trial
we attempt to fit all 61 trial time delays, for a grand total of $2.4\times10^{11}$~trials.  We discard all solutions in which 
$\chi^2/N_{dof}>5.0$, since they contribute negligibly to the Bayesian integrals and consume unnecessary disk space. An example of a set
of good fits to the observed flux ratios is shown in Figure~\ref{fig:fits}. 

We use Bayes' theorem to marginalize over the results from our Monte Carlo simulation such that the probability of a given time delay $\Delta t$
\begin{equation}
   P(\Delta t| D) \propto \int P(D| \vec{p}, \Delta t) P(\vec{p}) P(\Delta t) d\vec{p},
\label{eq:bayes}
\end{equation}
where $P(D|\vec{p}, \Delta t)$ is the probability of fitting the observations in a given trial, $P(\vec{p})$ sets the priors on the microlensing
variables \citep[see][]{Kochanek2004,Kochanek2007} and $P(\Delta t)$ is the (uniform) prior on the time delay.  

The probability density for the time delay is displayed in Figure~\ref{fig:td}.  To first order, our analysis appears to have failed, since it did not converge on a
robust, single-valued estimate for the time delay.  We did succeed, however, in effectively eliminating the possibility of positive values for the delay $\Delta t_{AB}$, which is 
an improvement over our earlier results in \citet{Morgan2008}. Nonetheless, the present probability density for the delay is (at least) double-valued. 
We also attempted a traditional polynomial fit time delay analysis \citep[e.g.][]{Kochanek2006}, in which we fit the quasar's intrinsic and microlensing
variability using Legendre polynomials.  The best-fit results from this attempt, based on a 30$^{{\rm th}}$ order fit to the intrinsic and 3$^{{\rm rd}}$ order fit to the
microlensing variability, are also displayed in Figure~\ref{fig:td}.  Again, as was the case with the Monte Carlo technique, our polynomial fits did not converge on a single
value for the delay, in fact the (inferior) polynomial fitting technique still allows for some significant probability of a positive (B leading A) time delay.
Nevertheless, the Monte Carlo procedure allowed us to study the structure of the quasar with proper statistical weighting of the uncertainties created by the time delay, 
ultimately enabling a successful joint fit to the observed X-ray and optical lightcurves using our Monte Carlo method.  

\begin{figure*}
\epsscale{1.0}
\plotone{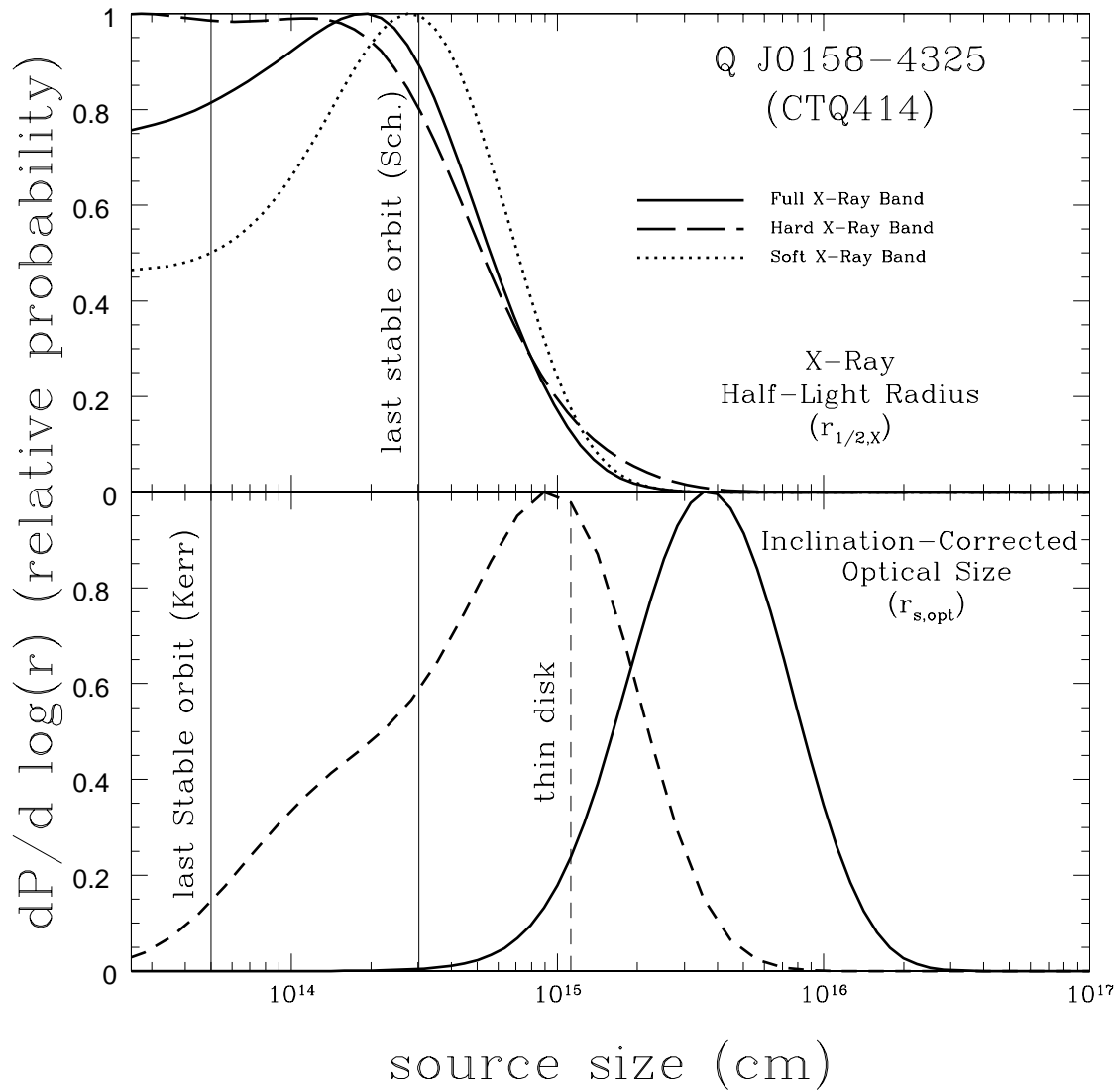}
\caption{Joint probability distributions for the X-ray half-light radius $r_{1/2,X}$ (Top Panel) and 
inclination-corrected optical source size $r_s$  (Bottom Panel) of Q~J0158--4325,
assuming $\cos i = 0.5$. In the X-ray panel, the solid, dotted, and dashed curves show the probability densities for the half light radius of the full band (rest
frame 0.9-18.4~keV), soft band (rest frame 0.9-3.0~keV) and hard band (rest frame 3.0-18.4~keV) continuum source, respectively. 
In the optical panel, the solid curve shows the estimate for $r_s$ from the simultaneous analysis of the X-ray and optical light curves.  
The dashed curve displays our original estimate of $r_s$ from \citet{Morgan2008} based on only 4 seasons of optical monitoring data. 
The vertical lines show the last stable orbit in the Schwarzschild metric at $6 r_g$ and at $1 r_g$ for a maximally rotating black hole in the Kerr metric. 
The black hole mass $M_{BH}=1.6 \times 10^8 M_\sun$ was estimated by \citet{Peng2006} 
using the \ion{Mg}{2} emission line width.}
\label{fig:rs}
\end{figure*}

For the simultaneous optical and X-ray analysis, we employed the technique detailed in \citet{Dai2010}, 
in which a large number of good fits to the optical light curves are generated first.  To enable better resolution at smaller angular scales, we used a set of
$8192\times8192$ magnification patterns with an outer scale of $10 R_{E}$, providing a source plane pixel scale of 
$4.2 \times 10^{13}  (\langle M \rangle/ 0.3 M_\sun)^{1/2}$~cm, $\sim2$ times larger than the 
black hole's gravitational radius at $r_g = 2.5 \times 10^{13}$~cm. We discuss the limitations of this pixel scale in \S~\ref{sec:results}.  We also tested the Monte Carlo
simulation using the original $20 R_E$ magnification patterns to confirm that the $10 R_E$ patterns from the low $f_{M/L}$ mass models did not introduce any 
significant systematic errors.
We saved all physical variables from the good optical fits (e.g. trajectory, effective transverse 
velocity $v_e$) and then fit the X-ray light curves using the same path across the magnification pattern but a different range of source sizes. An example of a good fit
from this analysis is also displayed in Fig.~\ref{fig:fits}.
We then analyze the combined X-ray and optical solutions to yield a set of joint probability densities 
for the variables of interest, namely the X-ray and optical source sizes.  Since the time delay
was unknown, we performed this analysis for a range of assumed delays, but the X-ray size was insensitive to the resulting minor changes in the optical solutions.  We
also tested our optical solutions for the possibility of $20-30\%$ line emission contamination from the quasar's un-microlensed broad line region, but this led to only modest
changes in the size estimates \citep[see][]{Morgan2010,Dai2010}. 

Similar to our analysis in \citet{Morgan2008}, the 
unknown time delay introduces short-timescale noise in the optical difference lightcurve that the Monte Carlo code 
can successfully reproduce using non-physical transverse velocities.  To eliminate this non-physicality in our models, we applied a prior on the effective transverse velocity 
$10 \: \langle M/M_\sun \rangle^{1/2} {\rm \, km \, {s^{-1}}} \leq {\hat v}_e \leq 10^4 \: \langle M/M_\sun \rangle^{1/2} {\rm \, km \, s^{-1}}$ where ${\hat v}_e$ is in 
Einstein units.  To convert all variables from Einstein units of $\langle M/M_\sun \rangle^{1/2}$ into physical units, we combined the resulting probability density
for the effective transverse velocity $P({\hat v}_e)$ with the distribution for the true transverse velocity $P(v_e)$. We generated the physical model for $P(v_e)$
following the method of \citet{Kochanek2004}, where we estimated the peculiar velocities of the source and lens using the power-law fits of \citet{Mosquera&Kochanek2011}.
We found the transverse velocity of the observer by projection onto to the CMB dipole $v_{CMB}=328.7 \: {\rm km \, s^{-1}}$, and we 
used  $\sigma_{pec,l}= 277 \: {\rm km \, s^{-1}}$ and  
$\sigma_{pec,s} = 248 \: {\rm km \, s^{-1}}$ for the peculiar velocities of lens and source, respectively.  We estimated 
the velocity dispersion in the lens galaxy  $\sigma_* = 203 \: {\rm km \, s^{-1}}$ using the monopole moment of a singular isothermal ellipsoid (SIE) lens model.

\section{Results: Optical and X-Ray Continuum Sizes and Lens Galaxy Properties}
\label{sec:results}
 
In Figure~\ref{fig:rs}, we display the probability densities for the size of the continuum emission region at optical and X-ray wavelengths.  In our microlensing 
simulations, we assume a simple thin accretion disk model where the disk radiates locally as a blackbody 
\citep[e.g.][]{Shakura1973}.  Since we expect the optical accretion disk to be many times larger than the black hole's gravitational radius $r_g =  G M_{BH} / c^2$, 
we assume that we can safely ignore the inner edge of the accretion disk and the associated 
relativistic corrections \citep[see][]{Morgan2010}.  With these assumptions, the surface brightness at 
rest wavelength $\lambda_{rest}$ is
\begin{equation}
f_{\nu}(r) = {2 h_p c \over \lambda_{rest}^3} 
\left[ \exp \left( {r \over r_{s}} \right)^{3/4}-1 \right]^{-1}
\label{eqn:disk1}
\end{equation}                                                                                   
where the scale length
\newpage
\begin{eqnarray}
r_{s} & = &\left[ {45 G \lambda_{rest}^4 M_{BH} \dot{M} \over 16 \pi^6 h_p c^2} \right]^{1/3} \\
& = & 9.7 \times 10^{15} \left( {\lambda_{rest} \over {\rm \micron}} \right)^{4/3} 
\left( {M_{BH} \over 10^9 {\rm M_{\sun}}} \right)^{2/3}
\left({L \over \eta L_E} \right)^{1/3} {\rm cm} \nonumber
\label{eqn:thindisk}
\end{eqnarray}
is the radius at which the disk temperature matches the wavelength, 
$k T_{\lambda_{rest}} = h_p c/ \lambda_{rest}$, 
$h_p$ is the Planck constant, $k$ is the Boltzmann constant, $M_{BH}$ is the black hole mass,   
$\dot{M}$ is the mass accretion rate, $L/L_E$ 
is the luminosity in units of the Eddington luminosity, and $\eta=L/(\dot{M}c^2)$ is the accretion efficiency.   In previous investigations \citep[e.g.][]{Morgan2010,Poindexter2008}
we found accretion disk temperature profiles consistent with the $T(r) \propto r^{-3/4}$ predictions of thin disk theory, 
so we report the size of the optical continuum emission region in terms of this model.

\begin{figure}
\epsscale{1.0}
\plotone{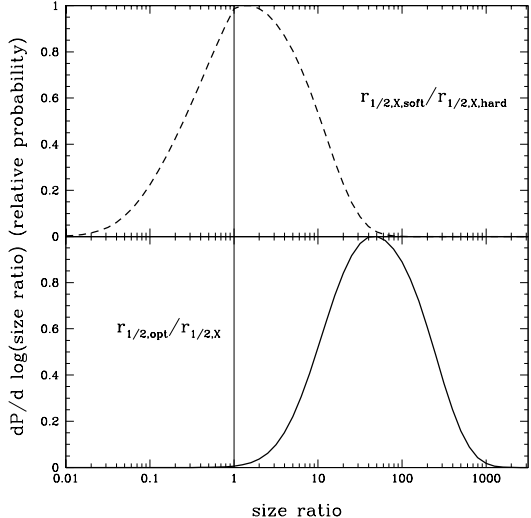}
\caption{Top Panel: Probability density for the ratio of the size of the soft X-ray continuum emission region to the size of the hard X-ray emission region.  
The expectation value $\log(r_{1/2,X,soft}/r_{1/2,X,hard}) = 0.2^{+0.6}_{-0.7}$ is consistent with unity, but tends toward a larger soft X-ray source.  
The solid vertical line at a size ratio of unity is plotted to guide the eye.
Bottom Panel: Probability density for the ratio of the optical half-light radius $r_{1/2,opt}$ to the half light radius of the full-band X-ray 
source $r_{1/2,X}$. The most likely ratio is $\log(r_{1/2,opt}/r_{1/2,X}) = 1.7 \pm 0.5$. }
\label{fig:sizeratios}
\end{figure}

Microlensing observations can estimate the temperature slope, since for 
$T \propto r^{-\beta}$ the disk size should scale as $r_{\lambda} \propto \lambda^{1/\beta} = \lambda^{4/3}$ for standard thin disk theory. 
Although there is general agreement that hotter regions (i.e. shorter wavelengths) are more compact, observational estimates of the slope $\beta$ have significant uncertainties
and dispersion between studies
\citep[e.g.][]{Anguita2008,Eigenbrod2008,Poindexter2008,Bate2008,Floyd2009,Blackburne2011}.  Fortunately, \citet{Mortonson2005} 
determined that microlensing statistics are insensitive to the choice of model surface brightness profile and depend primarily on the half-light radius.
Thus, the size we report here can easily be
compared to other estimates using the half-light radius $r_{1/2} = 2.44 r_{s}$.   Also, since microlensing statistics are actually sensitive to the projected area of the 
quasar's accretion disk, we must account for the unknown inclination angle $i$.  We assume that the optical emission comes from a geometrically thin disk so that the 
de-projected estimate of the size is $r_s = r_{s,obs}/(\cos i)^{1/2}$ for random orientations.  For simplicity, we use the same surface brightness distribution
in X-rays but assume that the emission is geometrically thick, so we report the half-light radius $r_{1/2,X}$.

In \citet{Morgan2010} we demonstrated that in the context of the thin accretion disk model, a quasar's 
specific luminosity $L_{\lambda}$ can also be used to estimate the size of 
its accretion disk at that rest-frame wavelength.  In the observed-frame $I$-band, for example, the so-called ``flux-based'' thin disk scale radius is
\begin{eqnarray}
r_{s,I} & = & 2.83 \times 10^{15} {1 \over \sqrt{\cos i}} \left( { D_{OS} \over r_H } \right) \nonumber \\
& & \times \left( { \lambda_{I,obs} \over {\rm \micron} } \right)^{3/2}
10^{-0.2(I-19)} \: h^{-1} \: {\rm cm}.
\label{eqn:fluxsize}
\end{eqnarray} 
There is now significant evidence in the literature that flux(luminosity)-based size estimates are consistently smaller than 
microlensing-based measurements \citep[e.g.][]{Poindexter2008,Hainline2012,Blackburne2012}, and
in \citet{Morgan2010}, we found this to be true in 10 out of the 11 quasars in our sample.  The one outlier in that sample was Q~J0158--4325, 
whose flux- and microlensing-based sizes were consistent. Integration of our updated $r_{s}$ probability density and correction for an average inclination
of $\cos i = 0.5$ yields an optical accretion disk size estimate of  $\log[(r_{s}/{\rm cm}) (\cos(i)/0.5)^{1/2}] = 15.6 \pm 0.3$ at $\lambda_{rest} = 0.31 {\rm \mu m}$,
the rest-frame center of the $R$-band. So in our new, more precise measurement, 
the flux-based size ($\log(r_{s,flux}/{\rm cm}) = 15.2\pm0.1$) is now $\sim0.4$~dex smaller than
the microlensing-based size, consistent with the discrepancies in the other systems. 

Probability densities for the half-light radii of the full, hard and soft X-ray continuum emission 
regions are also displayed in Figure~\ref{fig:rs}.  Since there is no simple spatial model for the 
X-ray continuum source, we report the X-ray sizes in terms of the half-light radius $r_{1/2,X}$ for X-ray 
emission at rest-frame energies 0.9-18.4~keV (full band), $0.9-3.0$~keV (soft band) and 3.0-18.4~keV (hard band).  
Integration of the probability density for the combined hard and soft X-ray flux yields an estimate for the 
full band half light radius $\log(r_{1/2,X}/{\rm cm}) = 14.2^{+0.5}_{-0.6}$. The soft and hard size estimates are consistent with that of the full band and each other, where
the half light radius for soft X-ray emission is
$\log(r_{1/2,X,soft}/{\rm cm}) = 14.3^{+0.4}_{-0.5}$ and the upper limit on the half-light radius for hard X-ray emission is $\log(r_{1/2,X,hard}/{\rm cm}) \leq 14.6$. 
Because the pixel scale in our magnification patterns is $7.6 \times 10^{13}$~cm for a 1~$M_{\sun}$ star, the lower limit on our X-ray size estimates is not fully
resolved.  In essence, once a trial source becomes much smaller than the pixel scale, all solutions become equally likely.  In the case of the hard X-ray source, this
prevented any convergence on a lower limit, so we also compiled a probability density for the
ratio between the soft and hard X-ray half-light radii $\log(r_{1/2,X,soft}/r_{1/2,X,hard}) = 0.2^{+0.6}_{-0.7}$.  This probability density is plotted in Figure~\ref{fig:sizeratios}
along with the ratio of the optical half-light radius to that in the full X-ray band, which peaks at $\log(r_{1/2,opt}/r_{1/2,X}) = 1.7 \pm 0.5$.

In other systems, \citet{Morgan2008,Dai2010} and \citet{Blackburne2012} have all found X-ray continuum regions that seem to be concentrated in the vicinity
of the black hole's last stable orbit. In Q~J0158--4325, the black hole mass $M_{BH} = 1.6 \times 10^8 \, {\rm M_{\sun}}$ \citep{Peng2006} implies a
gravitational radius $\log(r_{g}/{\rm cm}) = 13.4$ and last stable orbit $\log(r_{stable}/{\rm cm}) =14.2$ at $6 r_g$ in the Schwarzschild metric. 
Consistent with our results in HE~0435--1223, RXJ~1131--1231 and
PG1115+080, we find that the majority of the X-ray emission in Q~J0158--4325 emerges from a region in the vicinity of the last stable orbit. We also
find that the soft and hard X-ray continuum emission regions seem to be the same size, consistent with the findings of \citet{Blackburne2012} for HE~0435--1223. 

\begin{figure}
\epsscale{1.1}
\plotone{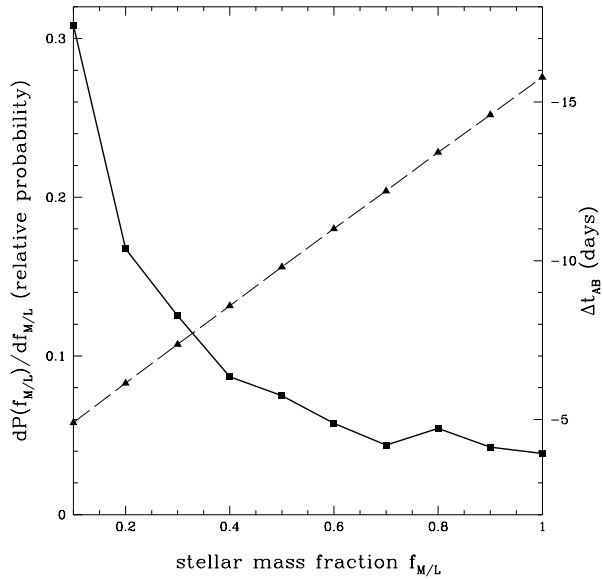}
\caption{Probability distribution for $f_{M/L}$, the fraction of the lens galaxy
mass in the constant M/L ratio (de Vaucouleurs) component (solid curve with square points).  The time delay values
predicted by each lens model are plotted with triangles and connected by a dashed curve.  These points correspond to the time delay scale on the right axis.
A modest trend favoring dark-matter dominated models is visible. 
\label{fig:fml}}
\end{figure}

In Figure~\ref{fig:fml}, we display the results of marginalizing the fits from our Monte Carlo simulation over all variables other than the stellar mass fraction in the macroscopic
mass model.  The resulting constraint on $f_{M/L}$ favors a dark-matter dominated halo, which seems to be marginally consistent with our 
expectations for a time delay of approximately $-8$~days.

\section{Discussion and Conclusion}  
 \label{sec:conclusions}

To demonstrate the improvement in accuracy gained by adding four seasons of new optical monitoring data and simultaneous analysis with X-ray lightcurves, 
we also show the size distribution from our first analysis of this system \citep{Morgan2008} in Fig.~\ref{fig:rs}. 
While our new measurement is statistically consistent with our previous estimate, the new expectation value is 
nearly 0.6~dex larger than our original analysis and has nearly twice the precision.  We attribute this
difference to two effects. First, while we failed to measure the time delay, the new data substantially reduced the allowed range to
$-30$~days~$\leq \Delta t_{AB} \leq 0$~days, leading to a significant narrowing of the source size probability density.
Flux ratios estimated for an incorrect delay have extra power on short time scales which the model tries to fit using a more compact source size, leading to a 
broadening of the source size distribution toward smaller sizes. 
Since the time delay signal with our four new seasons of monitoring data has effectively ruled out positive delays,
there are fewer cases in which the Monte Carlo code successfully reproduced the noise from an incorrect delay
using a small accretion disk size.  Second, the simultaneous analysis
of the (admittedly) sparse X-ray light curves provides a very strong constraint on the effective velocity distribution $v_e$, which, in turn, constrains $r_s$.
\citet{Blackburne2012} gained similar statistical leverage from the use of X-ray photometry in their analysis.

We have made significant improvements in our understanding of the QJ0158--4325 continuum source, but we are still unable to measure the time delay.  This may be due to 
inadequacies in our data or techniques, such as uncertainties in our
photometric measurements and/or insufficient signal from the time delay compared to the signal from microlensing, or it could possibly be due to our oversimplification
of the accretion disk model. For example, \citet{Abolmasov2012} demonstrate that general relativistic effects become important during caustic crossing events and 
can explain high-magnification features in the light curves of SBS~1520+530 and Q~2237+0305.  Several such features are present in the light curves of
Q~J0158--4325 and are not perfectly fit by our present analysis.  To be fair, the results of \citet{Abolmasov2012} require disks with large inclination
angles, and we see no evidence for this in Q~J0158--4325.  Nevertheless, we are left to conclude that the modeling of high-amplitude microlensing events
may be improved by the inclusion of a fully relativistic disk model or a more sophisticated model for the motions of stars in the lens galaxy \citep[e.g.][]{Poindexter2010}.

At X-ray wavelengths, a consistent picture of the structure of the quasar continuum emission region is beginning to emerge.  In the four systems with quantitatively
measured continuum emission regions, HE~0435--1223 \citep{Blackburne2012}, RXJ~1131--1231 \citep{Dai2010},  PG~1115+080 \citep{Morgan2008} and 
now QJ~0158--4325, the X-ray continuum emission source is located at $\sim 10 \, r_g$.  
Improving on our previous work, we performed
the second complete analysis of the hard and soft X-ray variability, leading to independent estimates for the sizes of the soft and hard X-ray sources.  While our hard and soft size
estimates are formally consistent with each other, there are some indications that the hard X-ray source is smaller than the soft source. 
As we discussed in \citet{Chen2012}, these data were not designed to produce good statistical estimates for the hard and soft band separately, so we expect that adding
more X-ray epochs with better photon statistics should lead to markedly improved X-ray size estimates.

\acknowledgements
This material is based upon work supported by the National Science Foundation under Grant No. AST-0907848. C.W.M. also gratefully
acknowledges support from the Research Corporation for Science Advancement. C.W.M., L.J.H., C.S.K., J.A.B., A.M.M., X.D. and G.C. are supported by 
Chandra X-Ray Center award 11700501. 
M.T., F.C. and G.M. acknowledge support from the Swiss National Science Foundation. C.S.K., J.A.B. and A.M.M. are supported by NSF grant AST-1009756.

{\it Facilities:} \facility{CTIO:2MASS (ANDICAM)}, \facility{HST (NICMOS, ACS)}, 
\facility{Swiss 1.2m Telescope}, \facility{{\it Chandra}} (ACIS)

\end{document}